\documentclass[journal=apchd5,manuscript=article]{achemso}

\usepackage{chemformula} 
\usepackage[T1]{fontenc} 

\author{Alisson Rodrigues de Paula}
\author{Saïd Idlahcen}
\author{Denis Lebrun}
\author{Pierre-Henry Hanzard}
\author{Ammar Hideur}
\author{Thomas Godin}
\email{thomas.godin@coria.fr}
\affiliation[CORIA]{CORIA UMR 6614, CNRS - University of Rouen Normandie - INSA Rouen Normandie, Rouen, France}

\title[An \textsf{achemso} demo]
  {Versatile ultrafast single-shot imaging}

\abbreviations{IR,NMR,UV}

\begin{document}

\begin{abstract}
Ultrafast single-shot imaging techniques now reach frame rates of tens of tera-frame-per-second (Tfps) and long sequence depths but are often too complex for large-scale use, both in terms of image acquisition and reconstruction. We propose an extremely simple yet high-performance method that leverages the capabilities of two prominent technologies: acousto-optical pulse shaping and light-field based hyperspectral imaging. We demonstrate the capabilities of the technique by capturing laser-induced phenomena at frame rates on par with the state-of-the-art, and with the potential to reach the peta-frame-per-second, while keeping a versatile setup that is easily adaptable to various input pulse shapes and dynamic events. Furthermore, an extra degree of freedom is added to the system through the use of digital in-line holography on the single-shot motion pictures. The agility and performance of this technique could then open up new horizons for single-shot imaging techniques, making them accessible to a wider community.
\end{abstract}

\section{Introduction}

Ultrafast optical imaging is a ubiquitous tool in many areas, spanning from fundamental sciences to applied research, where the need to capture ultrashort dynamics and transient events on microscopic scales has fueled the development of a myriad of remarkable techniques. Sub-ps resolutions are now reached using time-resolved pump-probe methods \cite{Fischer2016invited, Xie2021situ}, but such techniques are, however, limited to the capture of highly reproducible events. Over the last decade, significant efforts have therefore been dedicated to the realization of single-shot imaging methods with tera-frame-per-second (Tfps) frame rates to record non-repetitive or irreversible phenomena \cite{Liang2018review,Yao2024capturing}.

On the one hand, self-luminescent events can be captured via passive detection systems, and among them compressed ultrafast photography (CUP)-based methods \cite{Qi2020single, Wang2020single,Lu2019compressed}, combining streak imaging with compressed sensing, stand out as they reach frame rates above 10 trillion fps. On the other hand, active detection systems are used when illumination is needed\cite{Zeng2023review}, and frame division has been obtained in different domains: spatial\cite{Yeola2018single}, angular\cite{Li2014single}, polarization\cite{Yue2017one}, spatial frequency\cite{Ehn2017frame} or spectral\cite{Nakagawa2014STAMP}. Currently, the highest frame rate - 219 Tfps with a sequence depth of 230 frames - has been achieved with a hybrid active-passive method \cite{Wang2023single}. Both families come with their own advantages and drawbacks, but every technique must compromise in one or several key aspects within the imaging chain: frame rate, sequence depth, exposure time, field of view, light throughput, image reconstruction, system complexity, and footprint. Most of the research has been focused on increasing both the imaging speed (fps) and sequence depth (number of frames acquired in a single shot), in general to the detriment of simplicity, thereby ruling out the use of single-shot techniques outside an optics laboratory.  Many of the above-mentioned methods then remain highly complex to implement and operate, and sometimes require heavy computational processing for image reconstruction. In addition, any adjustment of parameters such as temporal resolution or range often entails redesigning or recalibrating the whole experiment.

For the sake of simplicity, but without making concessions on performances, several approaches have been explored with reconstruction-free methods\cite{Zeng2023review}. Among them, sequentially timed all-optical mapping photography (STAMP)\cite{Nakagawa2014STAMP} proved efficient by encoding each frame into a unique frequency band of a chirped pulse, and by subsequently mapping the time-tagged images in the spatial domain. The two key stages of STAMP-inspired techniques, which have been the subject of much research \cite{Li2024sequentially,Saiki2020sequentially,Li20232d,Saiki2023single}, are: \textit{(i)} the spectro-temporal shaper for creating pulses with appropriate features, and \textit{(ii)} the image detection, which must efficiently separate distinct spectral frames.

For the spectro-temporal shaping, two strategies have recently proved relevant. The first one uses a pulse-stretching stage termed spectrum circuit\cite{Saiki2023singlecircuit}, particularly adapted to Mfps-to-Gfps frame rates, and based on free-space angular-chirp-enhanced delay\cite{Wu2017ultrafast} to provide pulse trains on different timescales without sacrificing exposure time.  The second one consists of using an acousto-optic programmable dispersive filter (AOPDF\cite{Tournois1997acousto,Maksimenka2010direct}) to generate a discrete train of sub-pulses for imaging\cite{Touil2022acousto}. The AOPDF mainly consists in a birefringent crystal in which the optical wave is tailored via an interaction with a RF-controlled acoustic wave. This AOPDF-based method, on which we will build on in this study, allows to easily and independently control the chirp (temporal range), frame rate, exposure time and number of sub-pulses, along with their relative amplitudes and phases. 

For the image detection (\textit{i.e.} recording of spectral frames), various solutions have been implemented to simplify the complex mapping device from the original STAMP \cite{Nakagawa2014STAMP}. Spectral filtering \cite{Suzuki2015sequentially,Suzuki2017single} techniques separate the frames by combining a filter with a diffractive optical element but are, however, highly sensitive to angular alignment and result in a significant loss in light throughput. Another option is to combine a microlens array with a diffraction grating\cite{Nemoto2020single}, as it is easily adaptable to different illumination sources. A promising alternative to these solutions is to replace the whole detection stage with a snapshot hyperspectral camera (HSC)\cite{Yao2021single}, which offers higher spatial resolution, good pixel efficiency, and small footprint. Nevertheless, the performances strongly depend on the HSC technology, and the maximum sequence depth and temporal resolution are intrinsically limited by the HSC's number of channels along with their individual bandwidths.

In this study, we propose a versatile ultrafast single-shot imaging (VERSUS) technique that builds upon the most prominent strategies for both pulse shaping and image detection: AOPDF-based spectro-temporal control and plenoptic (\textit{i.e.} light field) hyperspectral imaging. In particular, we use a custom-designed hyperspectral camera based on light-field technology together with tailored ultrashort pulses within a very simple and flexible experimental setup. With this system, we report ultrafast imaging with record-high adaptable frame rates and the potential to reach the peta-frame-per-second (Pfps). In addition, further agility is added by using digital in-line holography to easily reconstruct images at different planes along the imaging axis and providing access to phase-sensitive imaging. The VERSUS method is assessed by capturing laser-induced filamentation in air and plasma dynamics, and the precise frame interval control down to the few-fs is highlighted with several experimental test cases. This proof of concept study demonstrates the potential of the VERSUS technique to bring single-shot imaging outside of the controlled environment of photonics laboratories.
.

\section{Experiments, results and discussion}

\subsection{Principle and experiment}

\begin{figure}
	\centering
  \includegraphics[width=1.0\textwidth]{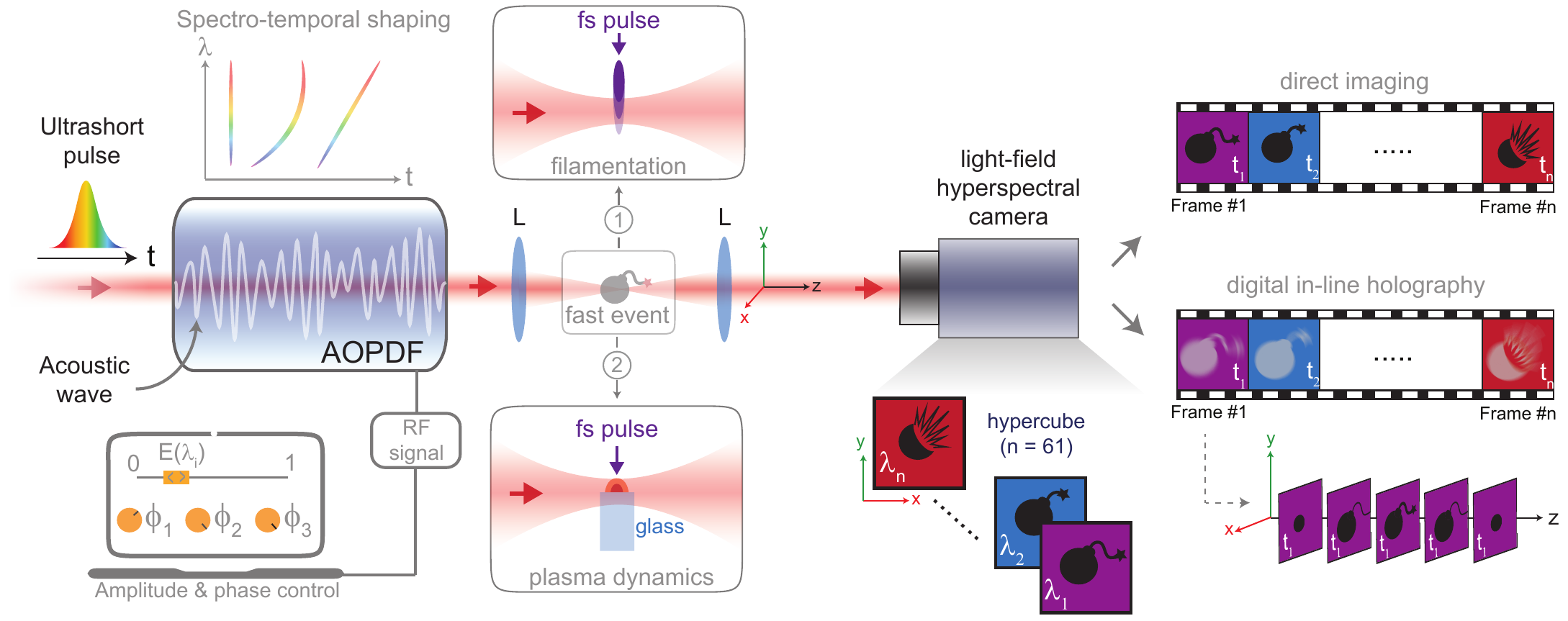}
  \caption{Principle of the versatile ultrafast single-shot (VERSUS) imaging technique and illustration of the laser-induced phenomena captured in this study. The spectro-temporal shape of ultrashort pulses is controled via the interaction with an acoustic wave within an acousto-optic programmable dispersive filter (AOPDF). The consecutive frames (i.e. spectral bands) are captured with a custom-designed hyperspectral camera based on light-field technology. Images can be retrieved directly for on-focus observation, or reconstructed at different z positions using digital in-line holography.}
  \label{setup}
\end{figure}
\figurename{ \ref{setup}} depicts the VERSUS imaging principle, leveraging AOPDF-based shaping and hyperspectral image capture. Ultrashort pulses from a Ti:Sapphire laser (800 nm, 150 fs) are first sent into a spectro-temporal tailoring stage, simply consisting of a RF-driven AOPDF, which carries out several tasks independently through the interaction of the optical wave with a controlled acoustic wave. Firstly, it imposes a dispersion profile suited to the events to be captured. At this stage, the dispersion parameters ($\phi_{n}$) can be finely adjusted up to higher orders. This allows to \textit{(i)} exactly adapt the frame rate to the event under study, in the tera-frame-per-second (Tfps) range, and \textit{(ii)} pre-compensate any chirp distortion of the input pulse, making the system particularly robust and adaptable. Secondly, the AOPDF performs amplitude equalization and then flattens the spectral profile to maintain a good and equivalent dynamic range in every spectral band. Pulses with on-demand continuous spectro-temporal profiles are then used to capture a dynamic scene (laser-induced phenomena in this study), and eventually each spectral sub-band (\textit{i.e.} each frame) is acquired using a custom-designed snapshot multi-aperture hyperspectral camera (HSC). This HSC, suited to the laser source, is based on light-field technology, enabling multi-dimensional imaging by fully exploiting all dimensions of the plenoptic function\cite{Ihrke2016principles, Wu2017light}. It provides remarkable spectral resolution over a broad bandwidth (1 nm resolution over a 60 nm bandwidth here, \textit{i.e.} 61 frames), enabling seamless integration with standard broadband fs lasers. As a comparison, in other ultrafast imaging strategies employing hyperspectral acquisition (\textit{e.g.} see Ref. \cite{Yao2021single}) the HSCs, often based on Fabry-Pérot filters atop standard sensors in a pixel-level mosaic layout, exhibit tens of nm wide channels and thereby require supercontinuum illumination sources, considerably increasing the complexity of the setup. The VERSUS technique (full description of the setup in Supporting Information) then achieves an extreme simplicity with only two main components to align, while reaching state-of-the-art performances with multi-tens of Tfps frame rates and 61 images, and the potential to reach the Pfps when imposing very small chirp values for the shaped pulses. Single-shot images can be recorded directly in focus but, in addition, out-of-focus images can also be used to recover depth (axial) information through simple digital in-line holography (DIH) tools, which allows image reconstruction at different planes along the optical axis, as demonstrated in a previous study \cite{Touil2022acousto}.   

\subsection{Capture of laser-induced ultrafast events}

\begin{figure}
	\centering
  \includegraphics[width=1.0\textwidth]{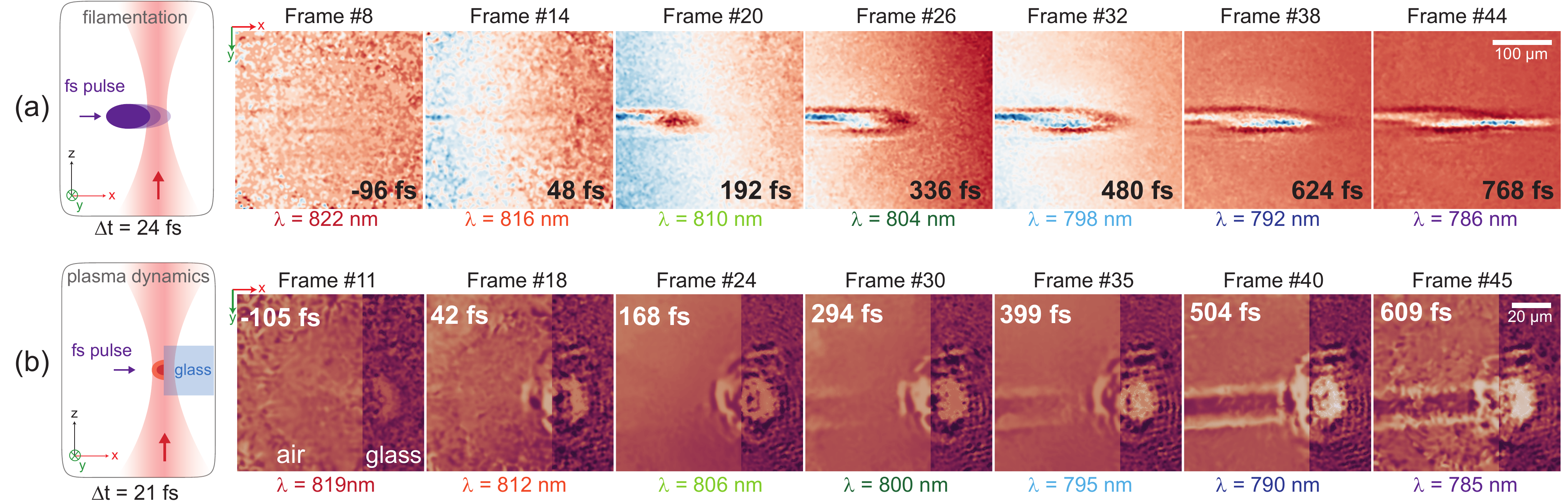}
  \caption{Tera-frame-per-second (Tfps) capture of laser-induced ultrafast phenomena in a single shot with the VERSUS technique: selection of 7 snapshots among the 61 spectral channels (i.e. frames) available. (a) Filamentation in air, with a 24 fs inter-frame interval. (b) Plasma generation on a glass target, with a 21 fs inter-frame interval. Videos 1 and 2 show the full sequences.}
  \label{testcases}
\end{figure}

The understanding of ultrafast light-matter interactions is crucial in various areas of physics and chemistry, such as plasma formation\cite{Hayasaki2017two} and radiation\cite{Zhang2018manipulation}, laser-induced microexplosions\cite{Rapp2015experimental} or laser micro- and nano-processing\cite{Pan2020ultrafast}, to cite a few. Such phenomena are then common testbeds for single-shot imaging techniques, as they require both high spatial and temporal resolutions. We validate the VERSUS technique by acquiring motion pictures of the dynamics of laser-induced phenomena, as previously depicted in \figurename{ \ref{setup}}. In our experiments, the temporal progression follows a sequence from higher to lower wavelengths as the probe pulses exhibit a positive chirp. Selected snapshots from the two experiments we performed are shown in \figurename{ \ref{testcases}}.  

We first tracked the filamentation of a unique femtosecond pulse in air, as shown in \figurename{ \ref{testcases}}(a). This process results from the dynamic competition between nonlinear self-focusing and plasma-induced defocusing and allows the propagation of the pulse within a plasma filament over distances well beyond the Rayleigh range of the beam \cite{Couairon2007femtosecond}. We used a frame rate of 41.7 Tfps (inter-frame interval of 24 fs) to record the creation of a plasma channel induced by focusing an ultrashort pulse (from the same source as the imaging pulse) with an energy of 1.1 mJ in air. The propagation is well captured and the raw images (minimal processing) extracted from the sequence (61 images in total) are presented in \figurename{ \ref{testcases}}(a).
We extracted physical information about the optical filament, including an average diameter of 18 µm and an average propagation velocity of 2.97$\times10^{8}$ m/s, which are consistent with parametric values reported in other studies \cite{Couairon2007femtosecond}.

In a second experiment, we studied plasma dynamics when focusing 200 µJ pulses (400 nm, 220 fs) in air and placing a glass sample in the beam path. The images obtained at a frame rate of 47.6 Tfps (inter-frame interval of 21 fs) are shown in \figurename{ \ref{testcases}}(b). Laser-induced phenomena (plasma creation and subsequent dynamics) that occur in the glass sample are clearly seen in the right part of the snapshots. This sequence also reveals that the latter occur prior (by a few hundred fs) to the appearance of the channel in air.
The complete sequence of both test cases are presented in videos 1 and 2. These two test cases demonstrate that the VERSUS imaging technique is on par with state-of-the-art single-shot methods in terms of performances (multi-10 Tfps and sequence depths >50), but with a considerably simpler design and without any reconstruction process, paving the way for direct use by non-specialists. 

\subsection{Spectral phase shaping for frame rate control}

To highlight the flexibility of the VERSUS technique and the gain in simplicity compared to other methods, we performed the filamentation experiment with different pulse tailorings. Although STAMP-based methods appear straightforward, most require selecting whether the train of sub-pulses should be uniformly spaced in the spectral or temporal domain, as the temporal or spectral mapping stages impose angle-frequency dependencies that are often nonlinear. For instance, the spectral band-pass filters in SF-STAMP\cite{Suzuki2015sequentially}, the diffraction gratings in SM-STAMP\cite{Saiki2020sequentially}, and the FACED mirrors in LA-STAMP\cite{Nemoto2020single} all have a sine dependence on the angle of incidence. In our previous demonstration combining SF-STAMP and AOPDF\cite{Touil2022acousto}, the spectral mapping also introduced a bottleneck in the spectro-temporal shaping of the AOPDF (temporal mapping). Operation was then only enabled within frequencies that could be detected simultaneously, leading to several trade-offs: \textit{(i)} mostly use the edges of the spectrum, where intensities are lower, thereby reducing the overall dynamic range, \textit{(ii)} use only a portion of the available bandwidth, still introducing cross-talk, or \textit{(iii)} exploit most of the bandwidth, at the expense of increased cross-talk and reduced spatial resolution.

Here, with the multi-aperture HSC and the suppression of the spectral bottleneck in the detection stage, we can thereby fully leverage the AOPDF’s capabilities, by shaping the entire spectro-temporal profile to achieve ultra-high frame rates that could theoretically reach the quadrillions of frames per second (Pfps). To this end, we accounted for all phase perturbations (initial chirp and subsequent dispersion) in the system prior to the sample plane. The AOPDF is then used to compensate for these phase distortions, including high-order dispersion terms. The AOPDF itself introduces a small spatial lateral chromatism (<$10^{-2}$ $\mu$m/nm) in its output, which can amplify third-order and higher-order dispersion effects when passing through thick lenses. These distortions can be pre-compensated before the detection stage. For fine corrections, a phase characterization device (\textit{e.g.} FROG) can be used to recover the phase immediately before the event, and to provide feedback to the AOPDF, ensuring optimal compensation.

\begin{figure}
	\centering
  \includegraphics[width=1.0\textwidth]{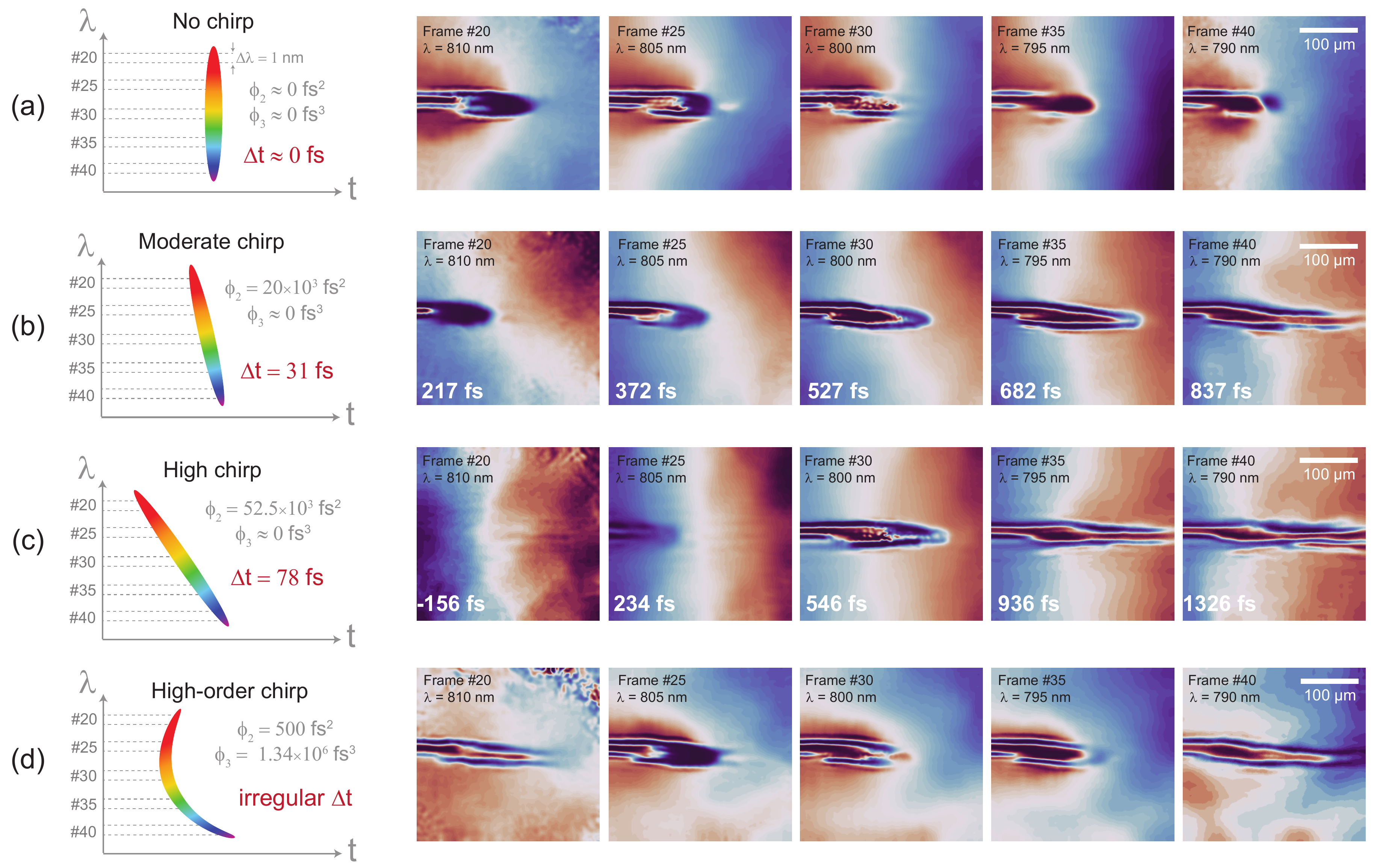}
  \caption{Capture of laser filamentation with different phase-shaped pulses. The phase values ($\phi_2$ and $\phi_3$) shown here correspond to their values after pulse pre-compensation. Linear chirp ($\phi_2$) is increased from (a) to (c), and a third-order phase ($\phi_3$) is added in (d). Complete sequences are available in videos 3, 4, 5 and 6.}
  \label{fpsControl}
\end{figure}

Once a probe pulse is fully unchirped (pre-compensated), one can then introduce a controlled chirp and adapt it to a given experiment. The user then has an easy and complete control over the system's temporal acquisition, including the frame rate and temporal window, as shown in \figurename{ \ref{fpsControl}}. A perfectly linear chirp, establishing a one-to-one correspondence between time and frequency (\textit{i.e.} each distinct frequency band directly maps a specific moment), is of course preferred for standard imaging conditions. Nonlinear chirps could however be used for monitoring accelerating/decelerating objects, or in single-shot multi-sampling frames for robust noise removal.

For instance, in \figurename{ \ref{fpsControl}}(a), the AOPDF was set to apply a second-order phase $\phi_2=-37.5\times10^{3}$ fs$^{2}$, a third-order phase $\phi_3=-1.35\times10^{6}$ fs$^{3}$, and a fourth-order phase $\phi_4= -1.03\times10^{3}$ fs$^{4}$, effectively counterbalancing dispersion and yielding an almost completely unchirped pulse. In \figurename{ \ref{fpsControl}}(b), after using the AOPDF to compensate the higher orders of dispersion in the same order of magnitude as in the previous case, the second-order phase was compensated by -17.5$\times10^{3}$ fs$^{2}$, resulting in an effective $\phi_2=20\times10^{3}$ fs$^{2}$ at the sample plane. Conversely, in \figurename{ \ref{fpsControl}}(c), instead of compensating dispersion, an additional second-order phase of 15$\times10^{3}$ fs$^{2}$ was introduced, leading to an effective $\phi_2= 52.5\times10^{3}$ fs$^{2}$.

As a result, \figurename{ \ref{fpsControl}}(a) displays a virtually static image (videos 3a and 3b), while in \figurename{ \ref{fpsControl}}(b), the propagation of the laser-induced filament was fully captured at 32.3 Tfps (31 fs inter-frame interval), maximizing the visualization of the event's evolution with an increased number of frames over the window of event. For instance, 32 frames were required in that case to observe the plasma fully crossing the field of view (FOV) (video 4). In contrast, in \figurename{ \ref{fpsControl}}(c), the increased chirp resulted in a lower frame rate of 12.8 Tfps (78 fs inter-frame interval), allowing to cover a larger temporal window, but with only 10 frames to capture the filament crossing the FOV (video 5). Finally, in \figurename{ \ref{fpsControl}}(d), no high-order dispersion was compensated, only most of the linear dispersion, resulting in an apparent forward-and-backward propagation in time (video 6). 

\subsection{Refocusing with digital in-line holography}

Standard in-focus imaging methods produce two-dimensional intensity-based images, such as those shown in \figurename{ \ref{testcases} and \ref{fpsControl}}, and are insensitive to phase variations. Recording out-of-focus interference patterns (e.g. when shifting the sample, lens or camera positions) can prove insightful when using digital in-line holography (DIH) for reconstructing the complex amplitude in real time, as further discussed in the Supporting Information. Here, we thereby used DIH to access the real and imaginary components of the optical field, and to extract not only the modulus but also the phase information. Phase reconstruction, in particular, is extremely useful for visualizing low contrast events in traditional intensity-based imaging\cite{Li2024sequentially, Xu2025single}. DIH then offers valuable three-dimensional insights, enabling the reconstruction of 2D images at different axial distances. This capability was previously leveraged to accurately determine the axial coordinate of the object of interest\cite{Touil2022acousto,Noury2023phase}. In addition, since wavelength is a simple parameter in DIH reconstruction, this method is particularly well suited to STAMP-based imaging. Here, we expand the use of this tool with a new approach to solve a common trade-off problem of axial resolution in microscopy. 

When observing microscopic-scale events, the magnifying optics used for visualization usually limit the depth of field. Therefore, if the sample is not perfectly aligned perpendicular to the lens, the observer must choose between what to keep in focus or the desired level of magnification. This trade-off is completely circumvented by applying DIH, which enables refocusing at different depths within a single magnified acquisition \cite{Lebrun_2003}.
Moreover, axial information allows retrieving angular information from the sample, thereby not restricting the capture to only flat or perfectly aligned events. 

\begin{figure}
	\centering
  \includegraphics[width=1.0\textwidth]{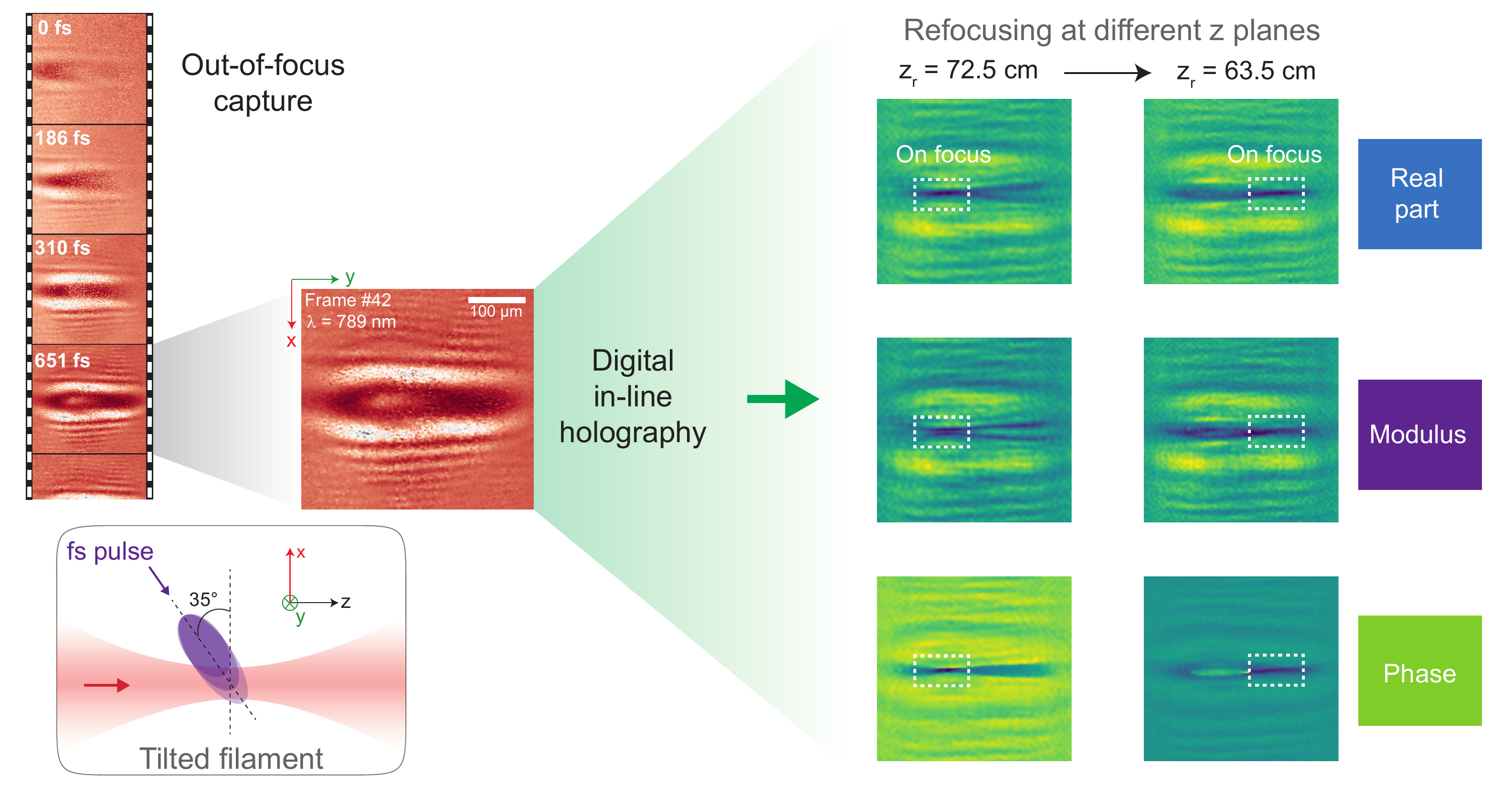}
  \caption{Observation of a laser-induced filament tilted relative to the imaging axis with digital in-line holography (DIH). Out-of-focus images (i.e. holograms) are captured in each spectral band and refocused at different positions along the optical axis using DIH. An example of reconstruction (imaginary part not shown here) on two distinct axial planes for one frame is represented, where the filament appears in focus at different transverse positions, therefore indicating a tilt in the object. Videos 7, 8, 9, and 10 show the reconstruction process on the full depth of field.}
  \label{holo}
\end{figure}

To demonstrate the potential of DIH with VERSUS imaging, we generated again a plasma channel in air, but this time crossing the probe beam at an arbitrary angle, as schematically illustrated in \figurename{ \ref{holo}}. To capture out-of-focus images, we relatively displaced the sample away from the in-focus position by 3 mm. 
From this acquisition, we could apply the reconstruction algorithm for each frame, which uses the recorded holograms (\textit{i.e.} patterns caused by the interference of the direct beam and the one scattered by the sampled object) to refocus the event at different axial planes. In \figurename{ \ref{holo}}, we combined the reconstructed 2D spatial information ($x_r$, $y_r$) from each frame with the reconstructed axial distance ($z_r$) to achieve a 3D reconstruction. By analyzing the orthogonal views ($y_r, z_r$ and $x_r,z_r$), we could infer and identify the in-focus plane within the images, either from real, imaginary, modulus, or phase information. The real physical spatial information ($x$, $y$, $z$) was then calculated using the system's optical element parameters. From this analysis, we determined that the plasma channel propagated at an angle of 35$^{\circ}$ relative to the camera plane. A more detailed mathematical analysis is provided in the Supporting Information, as well as additional considerations on the use of DIH together with experimental parameters.

\subsection{Discussion on the system capabilities}

As demonstrated, VERSUS imaging allows choosing the desired frame rate and selecting the temporal range of interest, in addition to adjusting the intensity of specific channels to prevent saturation. These features are directly linked to the AOPDF shaping capabilities, as fully characterized in Ref. \cite{Grabielle2011manipulation}. The amplitude control is limited by the spectral resolution, which is inversely proportional to both the crystal length and the group birefringence of the material along the propagation direction of the incident optical wave ($\delta n_{g} = \delta n_{g_{e}} - \delta n_{g_{o}}$) \cite{Kaplan2002theory}, as described in Equation \eqref{spectral_resolution}:
\begin{equation}
    \delta\lambda_{res} \simeq \frac{0.8 \lambda_{o}^{2}}{\delta n_{g} L}
    \label{spectral_resolution}
\end{equation}
Here, given the crystal length $L = 25$ mm, a central wavelength $\lambda_{o} = 800$ nm and with $\delta n_{g} \simeq 0.103$, the resulting spectral resolution $\delta\lambda_{res}$ is approximately $0.2$ nm. This value corresponds to a pulse-shaping spectral resolution that is five times narrower than the spectral window of each channel in the hyperspectral camera, ensuring no limiting factor on the amplitude control. This feature is especially useful as it allows for an increase in input optical power, which increases intensity at the edges of the spectrum while keeping high-intensity wavelengths below a saturation threshold. In this way, it is possible to improve the dynamic range in the peripheral channels and thus increase the effective sequence depth.

However, the increase in the input optical intensity has an upper limit constrained by the AOPDF's crystal damage threshold (30 µJ/pulse, as detailed in the Supporting Information). Conversely, the lower limit depends on user-defined criteria, as it involves trade-offs. Reducing the input power compromises the dynamic range of marginal wavelength channels, thereby decreasing the system's effective sequence depth. This limitation can be mitigated by applying electronic gain to the camera's CMOS sensor. Here, a multiplicative gain up to 10$\times$ had a manageable impact on image quality with minimal post-processing, and enabled an effective sequence depth of 11 frames for an input energy of only 100 nJ/pulse. 

On the other side, the temporal aspects are controlled through phase tailoring. As shown in Eq. \eqref{maximum_T}, the maximum temporal shaping capacity (\textit{i.e.} the maximum programmable time delay) is directly dependent on the crystal length.
\begin{equation}
    \Delta\tau_{max} = \frac{\delta n_{g} L}{c}
    \label{maximum_T}
\end{equation}
Here, for a commercially-available crystal with $L = 25$ mm, the maximum delay is approximately $8.5$ ps, which limits the available frame rates to over a hundred Gfps. However, the minimum time delay does not impose any limitations on the system and can be determined using the following relation:
\begin{equation}
    \tau_{g_{min}} \cong \frac{2\pi}{\Delta\omega SNR}\sqrt{\frac{\delta\omega}{\Delta\omega}} 
    \cong \frac{\lambda_{o}^{2}}{c \Delta\lambda SNR}\sqrt{\frac{\delta\lambda}{\Delta\lambda}}
    \label{min_groupdelay}
\end{equation}
Considering a typical RF signal-to-noise ratio (SNR) of $10^{4}$, a spectral resolution of $\delta\lambda = 0.2$ nm, and the HSC spectral resolution of $\Delta\lambda = 1$ nm as a limitation factor, the minimum group delay per HSC-channel-bandwidth $\tau_{g_{min}}$ is $0.10$ fs. Without any constraints to the system, the spectro-temporal profile could then be finely adjusted with a precision of 0.10 fs for each 1 nm of bandwidth, corresponding to a remarkable frame rate of 10 Pfps. The system could actually achieve an even higher frame rate at the expense of the sequence depth. For instance, with a spectral resolution of $\Delta\lambda = 2$ nm, meaning that every two channels in the HSC will be identical (reducing the available sequence depth by half), the minimum step is $\tau_{g_{min}} = 34$ as. This corresponds to an equivalent frame rate of 29.6 Pfps, which could prove relevant when studying ultrafast events on even smaller spatial scales.

In addition to characterizing the inter-frame interval, it is crucial to determine the exposure time of each frame (\textit{i.e.} temporal resolution). In this setup, the minimum exposure time per frame is dictated by the duration of the Fourier transform-limited (fully unchirped) pulse, meaning that the temporal resolution is inherently dependent on the laser source and its spectral profile. With the laser used in this study, we achieved a temporal resolution of $\sim100$ fs. Although linearly chirping the pulse slightly increases the exposure time, this effect can be mitigated by designing an adapted acoustic mask. Here, considering a 800 nm source with a theoretical minimum duration of a few fs (single-cycle), frame rates in the sub-Pfps range are required for an effective blurless acquisition where frames are unique and do not overlap in time (\textit{i.e.} exposure time shorter than the inter-frame interval). Our temporal control with sub-fs precision is thereby not a limiting factor, ensuring operation at the maximum imaging speed. The latter could even be increased to the Pfps with higher temporal resolutions using lasers with shorter wavelengths, as shown in Eqs. \eqref{spectral_resolution} and \eqref{min_groupdelay}. Note that all the devices used in this study are for instance available in the visible and UV ranges.  

Finally, in terms of spatial resolution, the main limiting factor is not the diffraction limit, but rather the effective pixel size of the camera. Although the physical pixel size of the CMOS sensor is 3.45 $\mu$m, a light-field based camera captures smaller images multiple times across the sensor. These values are then combined to produce a single pixel in the rendered data cube, resulting here in a larger effective pixel size of $\sim$22.22 $\mu$m. Nevertheless, a spatial resolution of 161.3 line pairs per millimeter (lp/mm) is achieved, allowing us to resolve objects larger than 3.1 $\mu$m.

A characteristic trade-off in multi-aperture light-field based HSC exists between spectral and spatial resolution due to the division of the sensor area. Since our HSC is a prototype, it was expected that the initial design would not deliver optimal performance. However, identifying these limitations has allowed us to explore solutions to minimize such a trade-off. One solution to improve spatial resolution is to reduce the number of spectral channels, which would increase spatial sampling but reduce the system's sequence depth. A second and more advantageous solution, without compromising the system's cutting-edge performance, is to use a larger CMOS sensor. Here, we used a 5-megapixel sensor, but 24-megapixel alternatives are readily commercially available, for instance. With such an upgrade and assuming 20 physical spectral channels (approximately 29 interpolated narrow channels), it is possible to achieve a pixel resolution of 1 megapixel/channel, which is 13 times greater than in this study (290$\times$275 pixels). This would allow for a reduction in the effective pixel size, subsequently increasing the final spatial resolution. A larger CMOS sensor would also enhance the system's ability to capture more fringes, in particular high spatial frequencies, in the out-of-focus images, thereby improving the spatial resolution of the DIH reconstructed images. Thus, the trade-off between spectral resolution, spatial resolution, and sequence depth can be refined through camera customization and optimized to match the specific laser source used. 


\section{Conclusions and prospects}

In this proof-of-concept study, we achieved ultrafast single-shot imaging with a simple and agile – yet powerful – setup synthetizing AOPDF-based pulse shaping and custom-designed hyperspectral imaging. Based on these two prominent approaches, we demonstrated state-of-the-art adaptable frame rates in the Tfps range, with relatively long sequence depths (61 frames) and the ability of the technique to adapt on demand to a wide variety of laser sources and experimental conditions. Given the AOPDF shaping capabilities and the multi-aperture hyperspectral camera spectral resolution, peta-fps frame rates could be obtained, but with a limited exposure time. In addition, digital in-line holography can be used on out-of-focus time-tagged images to produce dynamic 3D reconstructions, and thereby to capture events at different axial positions within the field of view, while it also enables phase-sensitive acquisition. All these features together make the technique highly versatile, with reduced optical complexity and minimal alignment, therefore broadening the applicability of single-shot imaging techniques outside photonics laboratories.     
Finally, VERSUS imaging can still be improved and further simplified in its three main components. Firstly, regarding the AOPDF, the available temporal window (limited to 8.5 ps here) can be extended by using longer birefringent crystals, as some are already commercially available. To widen this window to the 100 ps-to-ns (Gfps) range while keeping short exposure times, our technique could be combined with other technologies such as the spectrum circuit \cite{Saiki2023single}. Secondly, regarding the hyperspectral camera, the sequence depth could easily be increased by using a larger CMOS sensor together with an adapted filter or a laser with a broader bandwidth. Lastly, a specifically designed laser source, such as a fiber laser with a tailored profile, could be implemented to minimize the system footprint, while a more stable and shorter-pulse laser could enhance temporal resolution.

\begin{acknowledgement}
The authors acknowledge financial support from the Agence Nationale de la Recherche Labex EMC3 (QuantyPHy project) and Région Normandie (HOPTIM project). The authors would like to thank Tristan Guezennec (CORIA), and the team of Cubert GmbH for fruitful discussions. 
\end{acknowledgement}

\begin{description}
\item[Disclosures] 
The authors declare no conflicts of interest.

\item[Data availability]
Data underlying the results presented in this paper are not publicly available at this time but may be obtained from the authors upon reasonable request.
\end{description}

\begin{suppinfo}

The following files are available for supporting content:
\begin{itemize}
    \item Detailed experimental setup, considerations about real-time detection and image reconstruction, and principles of digital in-line holography.  

    \item \textbf{Video 1:} Filamentation in air captured with a 24 fs inter-frame interval (41.67 Tfps). Images were post-processed using a non-local means denoising algorithm to \textit{enhance contrast}. This video serves as a supplement to \figurename{ \ref{testcases}}(a).

    \item \textbf{Video 2:} Plasma generation on a glass target captured with a 21 fs inter-frame interval (47.62 Tfps). Images were post-processed using a non-local means denoising algorithm to \textit{enhance contrast}. This video serves as a supplement to \figurename{ \ref{testcases}}(b).

    \item \textbf{Video 3a:} Filamentation in air captured with a complete unchirped probe pulse ($\sim$0 fs inter-frame interval). Images are \textit{raw data} from the camera with the background subtracted (without any further post-treatment). This video serves as a supplement to \figurename{ \ref{fpsControl}}(a).
    
    \item \textbf{Video 3b:} Filamentation in air captured with a complete unchirped probe pulse ($\sim$0 fs inter-frame interval). Images were post-processed using a non-local means denoising algorithm to \textit{enhance contrast}. This video serves as a supplement to \figurename{ \ref{fpsControl}}(a).
    
    \item \textbf{Video 4:} Filamentation in air captured with a 31 fs inter-frame interval (32.26 Tfps). Images were post-processed using a non-local means denoising algorithm to \textit{enhance contrast}. This video serves as a supplement to \figurename{ \ref{fpsControl}}(b).

    \item \textbf{Video 5:} Filamentation in air captured with a 78 fs inter-frame interval (12.82 Tfps). Images were post-processed using a non-local means denoising algorithm to \textit{enhance contrast}. This video serves as a supplement to \figurename{ \ref{fpsControl}}(c).

    \item \textbf{Video 6:} Filamentation in air captured with irregular inter-frame interval. Images were post-processed using a non-local means denoising algorithm to \textit{enhance contrast}. This video serves as a supplement to \figurename{ \ref{fpsControl}}(d).
    
    \item \textbf{Video 7:} \textit{Real-part} reconstruction of tilted filamentation in air's complex amplitude using digital in-line holography. The reconstruction was performed using raw camera data with background subtraction from an out-of-focus frame. \textit{No post-treatment} was applied to the reconstructed holographic images either. This video serves as a supplement to \figurename{ \ref{holo}}.
  
    \item \textbf{Video 8:} \textit{Imaginary-part} reconstruction of tilted filamentation in air's complex amplitude using digital in-line holography. The reconstruction was performed using raw camera data with background subtraction from an out-of-focus frame. \textit{No post-treatment} was applied to the reconstructed holographic images either. This video serves as a supplement to \figurename{ \ref{holo}}.
    
    \item \textbf{Video 9:} \textit{Modulus} reconstruction of tilted filamentation in air's complex amplitude using digital in-line holography. The reconstruction was performed using raw camera data with background subtraction from an out-of-focus frame. \textit{No post-treatment} was applied to the reconstructed holographic images either. This video serves as a supplement to \figurename{ \ref{holo}}.
    
    \item \textbf{Video 10:} \textit{Phase} reconstruction of tilted filamentation in air's complex amplitude using digital in-line holography. The reconstruction was performed using raw camera data with background subtraction from an out-of-focus frame. \textit{No post-treatment} was applied to the reconstructed holographic images either. This video serves as a supplement to \figurename{ \ref{holo}}.
\end{itemize}

\end{suppinfo}

\bibliography{biblio}

\end{document}